\documentclass[conference,10pt]{IEEEtran}
\IEEEoverridecommandlockouts
\usepackage{pifont}% http://ctan.org/pkg/pifont
\usepackage{tabu}
\usepackage{longtable}
\usepackage{graphicx}

\usepackage{dsfont}
\usepackage{multicol}
\usepackage{amsmath, amssymb, bm, cite, epsfig, psfrag, mathtools}
\usepackage{epstopdf}
\usepackage{changepage}
\usepackage{graphicx}
\usepackage{fancyhdr}
\usepackage{bbm}
\usepackage{algorithm,algorithmic}
\usepackage{array}
\usepackage[margin=10pt,font=small]{caption}
\usepackage{bbm}
\usepackage{multirow}
\usepackage[usenames,dvipsnames]{xcolor}

\usepackage{etoolbox}
\usepackage{pbox}
\usepackage[width=0.48\textwidth]{caption}
\usepackage{colortbl}
\usepackage{bm}
\usepackage{caption}
\usepackage{subcaption}
\usepackage{amsmath}
\usepackage{color}
\usepackage{enumitem}
\usepackage{datetime}
\usepackage{amsthm}
\usepackage{dsfont}
\usepackage{mathtools}
\usepackage[T1]{fontenc}
\usepackage{nicefrac}

\newtoggle{conference}
\togglefalse{conference} % Use for arxiv version -- also change documentclass
\usepackage[letterpaper,left=0.75in,right=0.75in,top=0.75in,bottom=0.75in,footskip=.25in]{geometry}
\def\beq{\begin{equation}}
\def\eeq{\end{equation}}
\def\beqa{\begin{eqnarray}}
\def\eeqa{\end{eqnarray}}
\def\beqan{\begin{eqnarray*}}
\def\eeqan{\end{eqnarray*}}

\def\EE{{\mathbb{E}}}

\def\x{\times}

\newtheorem{Remark}{Remark}
\newtheorem{proposition}{Proposition}

\newtheorem{theorem}{Theorem}

\def\tm1{t\! - \! 1}
\def\tp1{t\! + \! 1}

\def\gbf{\mathbf{g}}

\def\Ibf{\mathbf{I}}

\def\wbf{\mathbf{w}}

\def\xbf{\mathbf{x}}

\def\Var{{\rm Var}}

\usepackage{pifont}

\begin{document}

\bibliographystyle{IEEEtran}

% **************************************************************************************************
% TITLE PAGE
% **************************************************************************************************
\title{Sectoring in Multi-cell Massive MIMO Systems}

% \author{\IEEEauthorblockN{$\text{Shahram Shahsavari}^\dagger$, $\text{Parisa Hassanzadeh}^\dagger$, $\text{Alexei Ashikhmin}^\ddagger$, and $\text{Elza Erkip}^\dagger$}
% \IEEEauthorblockA{$\dagger$ Department of Electrical and Computer Engineering, NYU Tandon School of Engineering}
% \IEEEauthorblockA{$\ddagger$ Nokia Bell Laboratories}
% \thanks{}

% }

\author{Shahram Shahsavari, Parisa Hassanzadeh, Alexei Ashikhmin, and Elza Erkip
\thanks{S. Shahsavari, P. Hassanzadeh  and  E. Erkip are with the ECE Department of New York University, Brooklyn, NY. Email: \{shahram.shahsavari,ph990, elza\}@nyu.edu}
\thanks{A. Ashikhmin is with Bell Labs, Nokia, Murray Hill, NJ, USA. Email:  alexei.ashikhmin@nokia.com}
}

\maketitle
% ********************************************************************************************************
% ABSTRACT
% ********************************************************************************************************
\begin{abstract}
In this paper, the downlink of a typical massive MIMO system is studied when each base station is composed of three antenna arrays with directional antenna elements serving $120^\circ$ of the two-dimensional space. A lower bound for the achievable rate is provided. Furthermore, a power optimization problem is formulated and as a result, centralized and decentralized
power allocation schemes are proposed. The simulation results reveal that using directional antennas at base stations along with sectoring can lead to a notable increase in the achievable rates by increasing the received signal power and decreasing `pilot contamination' interference in multi-cell massive MIMO systems. Moreover, it is shown that using optimized power allocation can increase \emph{0.95-likely} rate in the system significantly.
\end{abstract}

% Sectorization is less expensive than cell-splitting, as it does not require the acquisition of new base station sites. Cell
% Sectoring It basically involves replacing an omni directional antenna at the base station by several directional antennas. Cell
% sectoring is done mainly to reduce factors such as co-channel interference. It offers the following advantages:
% > Better S/I ratio
% > Reduces interference, increases capacity
% > Reduces cluster size, more freedom in assigning channels
% Advantages of Cell Sectoring
% > Increased number of antennas per base station
% > Decrease in trunk efficiency
% > Loss of traffic
% > Increased number of handoffs
% Limitations of Cell Sectoring

\section{Introduction}\label{sec:introduction}
With the advent of new technologies such as smart phones, tablets, and new applications such as video conferencing and live streaming, there has been a dramatic increase in the demand for high data rates in cellular systems. On the other hand, it is challenging to achieve high enough data rates in the crowded sub-6 GHz spectrum. Multi-user Multi Input Multi Output systems with large number of antennas (known as massive MIMO), have shown a great potential to achieve very large spectral and energy efficiencies, which makes them a strong candidate for 5G mobile networks \cite{larsson2014massive}.

In massive MIMO systems, base stations are usually equipped with a large number of antennas serving much smaller number of users each of which has an omnidirectional antenna. It is shown in \cite{marzetta2006much} that with a simple Time Division Duplex (TDD) protocol, it is beneficial to increase the number of base station antennas in a single cell massive MIMO network. More specifically, it is shown that received signal power is proportional to number of antennas while interference plus noise power is not. However, as shown in \cite{marzetta2010noncooperative}, another type of inter-cellular interference, called `pilot contamination', appears in multi-cell massive MIMO networks.
Typically training sequences should be short, since channels between base station and users change fast. This forces one to use nonorthogonal training sequences in neighboring cells, which causes pilot contamination whose power is proportional to the number of antennas at the base stations. Consequently, Signal to Interference plus Noise Ratio (SINR) converges to a a bounded value as the number of antennas tends to infinity.

%In \cite{pcp1}, a Large Scale Fading Precoding (LSFP) is proposed in which base stations cooperate using only large-scale fading channel coefficients. Zero-forcing LSFP allows to cancel the interference caused by pilot contamination completely. However, it is not optimal unless the base stations have very large number of antennas (e.g. $10^6$) which makes pilot contamination the dominant part of the interference. In \cite{pcp2}, the authors propose an optimized LSFP which is able to achieve significant improvements in users' rates in every massive MIMO system. Overall the results in \cite{pcp1,pcp2} imply that interference is a limiting factor in practical massive MIMO systems with finite number of antennas at each base station, and it can be reduced using large-scale precoding which imposes cooperation overhead to the system.

Most literature on massive MIMO considers omnidirectional base station antennas. It is well-known that using directional antennas along with sectorized antenna arrays at each base station is one of the methods to increase SINR in conventional cellular networks \cite{andrews2007overcoming}. Reference \cite{mehmood2013large} indicates the potential of using directional antennas in massive MIMO systems; however, it does not provide any performance analysis. In this paper, we consider the sectorized setting, analyze the performance of a massive MIMO system with directional antennas at each base station, and provide a lower bound on the achievable downlink rate of the users as a function of large-scale fading coefficients. We formulate a tractable downlink power optimization problem and suggest a centralized scheme to find the optimal power allocation. To reduce the communication and computation overheads, we also provide a sub-optimal decentralized scheme. A numerical comparison between a massive MIMO system with omnidirectional antennas at each base station and one with directional antennas shows that using directional antennas can improve the performance significantly.
 To the best of our knowledge, this is the first detailed study of massive MIMO systems with directional antennas.
%As opposed to \cite{pcp1,pcp2}, we do not consider LSFP to combat the interference and leave it as a future work. Instead, we formulate a tractable downlink power optimization problem and suggest a centralized scheme to find the optimal power allocation.

%The rest of the paper is organized as follows. We start with system model description in Sec. \ref{sec:system model}. In Sec. \ref{sec: Downlink Analysis}, we provide the analysis of the downlink performance and show the numerical results in Sec. \ref{sec: Numerical Analysis}. Finally we conclude the paper with Sec. \ref{sec: Conclusion}.

%As proposed in \cite{pcp1}, zero-forcing precoding (based on the large-scale fading coefficients) is able to remove the effect of
%pilot contamination, which can be used to achieve infinitely growing SINR as $M\rightarrow \infty$. However, with finite $M$,
%zero-forcing precoding is sub-optimal as the two types of interference become comparable. For the finite regime $M$, \cite{pcp2}
%designs an optimal precoder to combat the both types of interference. In this paper, we study the effect of sectoring in reducing
%inter-cellular interference, and do not consider such precoding techniques.
%
%We next consider the different strategies that can be employed by the network for downlink power allocation, and compare the system performance with those given in \cite{pcp1}, in Sec~\ref{sec: Numerical Analysis}. {\BLUE do we?}

%say how in conventional SINR increases.
%cite \cite{marzetta2010noncooperative}
%\newpage
\section{System Model}\label{sec:system model}
\begin{figure}[h]
  \centering
  \includegraphics[width=0.5\linewidth]{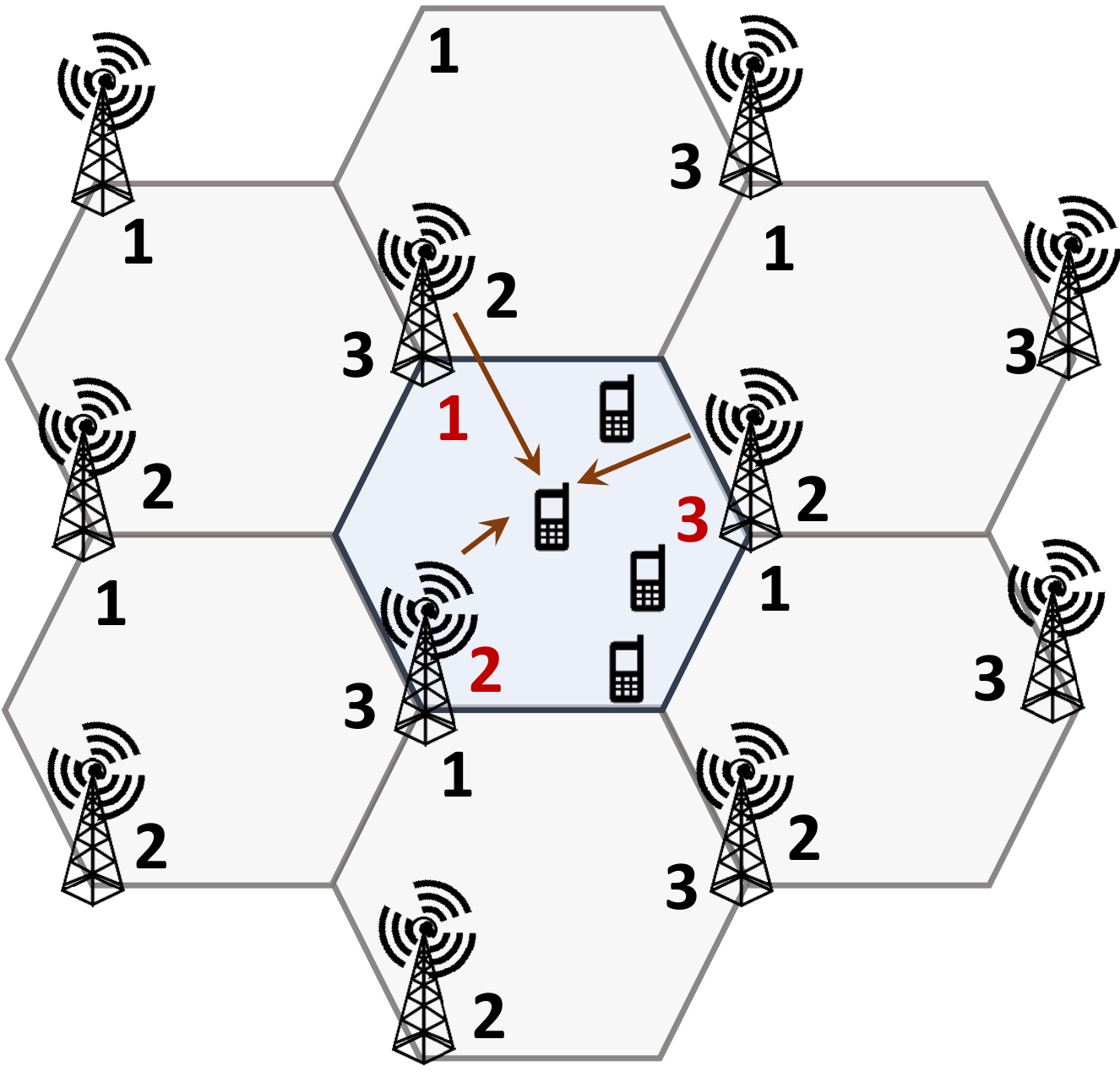}%{Figures/SecModel.pdf}
  \caption{Sectorized cellular system model}
  \label{fig: Sectored Model}
\end{figure}

We consider a two-dimensional sectorized hexagonal cellular network with TDD operation, composed of $L$ cells, each with $K$ mobile users. Cell sectoring is done such that three base stations are located at the non-adjacent corners of each cell, as shown in Fig.
\ref{fig: Sectored Model}, and each base station is equipped with three directional ($120^{\circ}$) antenna arrays such that each array
serves one of the three neighboring cells. As depicted in Fig.~\ref{fig: Sectored Model}, the users in each cell are served by the three antenna arrays that belong
to the base stations located on the corners of the cell. We assume that each directional array has $M$
directional antenna elements, hence there are $M_{B} = 3M$ elements at each base station. We also assume that users have single omnidirectional antennas.

In the following we denote cell $j$ by $C_j$, and user $k$ in cell $j$ by $U_{kj}$, where $j \in [L]$\footnote{We denote by $[N]$
the set of integers from 1 to $N$.} and $k\in [K]$. Each antenna array is uniquely identified by a cell-array index pair
$(j,i)$, and is denoted by $A_{ji}$, where $i \in [3]$ indicates the array located in corner $i$ of cell $j\in [L]$ (see Fig.~\ref{fig: Sectored Model}). %{$j \in [L]$ is the cell index served by the AA, and $i \in [3]$ indicates the array index in cell $j$(Fig. \ref{fig: Sectored Model}).}
User $U_{kj}$ communicates with all three arrays $A_{j1}$, $A_{j2}$, and $A_{j3}$ for uplink and downlink transmissions.

\subsection{Directional Antenna Model}\label{sec: antenna model}
\begin{figure}[h]
  \centering
  \includegraphics[width=0.95\linewidth]{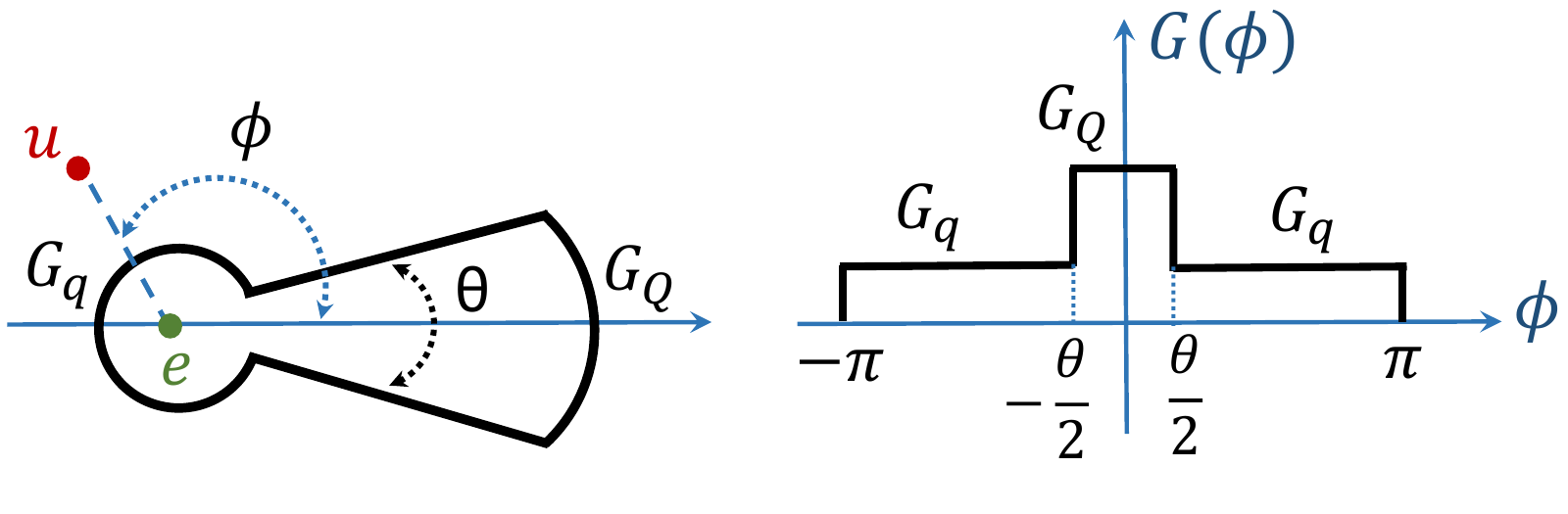}
  \caption{Simplified directional antenna power gain pattern}
  \label{fig:pattern}
\end{figure}

We adopt the simplified directional antenna model introduced in \cite{ramanathan2001performance}. Fig. \ref{fig:pattern} depicts the directivity (power gain) pattern of each array element, where $G_Q$ and $G_q$ are the main lobe and back lobe power gains, respectively, and $\theta$, chosen as
$2\pi/3$, is the beamwidth of the main lobe. Let $\phi\in [0,\pi]$ denote the angular position of a user placed at an angle $\phi$ relative to the
boresight direction of an antenna element, as in Fig.~\ref{fig:pattern}~(left), then the signal transmitted to and received from the user is multiplied by a gain equal to $\sqrt{G(\phi)}$.
%The signal transmitted to the user at $u$, denoted by $s_X$, is received by $X$ as $\sqrt{G(\phi_{XY})}g_{YX}s_X$ , where $g_{YX}$ denotes the channel coefficient between $Y$ and $X$.
%\vspace{-5pt}

Let $G_{ji}^{[kl]}$ denote the power gain between $U_{kl}$ and $A_{ji}$.
%More specifically, if $\phi^{[kl]}_{ji}$ denotes the angular position of $U_{kl}$ relative to the boresight direction of $A_{ji}$, then the power gain
%between $U_{kl}$ and $A_{ji}$, denoted by $G_{ji}^{[kl]}$, is
%\begin{equation}
%G_{ji}^{[kl]} =  \left\{
%\begin{array}{ll}
%   G_Q            & \mbox{if }   |\phi^{[kl]}_{ji}| \leq \frac{\theta}{2}, \\%\frac{\theta}{2}=60^{\circ}, \\
%   \\
%   G_q            & \mbox{if }   |\phi^{[kl]}_{ji}| > \frac{\theta}{2}. %\frac{\theta}{2}=60^{\circ}.
%   \end{array}
%   \right.
%   \label{antenna-pattern}
%\end{equation}
Note that all users in cell $C_j$, $j\in[L]$ are in the main lobe coverage of arrays
$\{A_{ji}: i\in[3]\}$, and therefore, observe the power gain $G_Q$ for any $k\in[K]$ and any $i \in [3]$. We assume a lossless antenna model
which implies that $G_Q+2G_q=3$, and $G_Q \leq 3$ due to the conservation of power radiated in all directions~\cite{ramanathan2001performance}.

\subsection{Channel Model}\label{subsec:channel model}
Due to TDD operation and channel reciprocity, downlink and uplink transmissions propagate similarly. We assume narrow-band flat fading channel model in which, the complex channel (propagation) coefficient between the $m$-th antenna element of $A_{ji}$ and $U_{kl}$ is
given by
\begin{align}
g^{[kl]}_{mji} = \sqrt {\beta^{[kl]}_{ji}} h^{[kl]}_{mji}, \label{full channel}
\end{align}
where $\beta^{[kl]}_{ji} \in \mathbb{R}^{+}$ is the large-scale fading coefficient, which depends on the shadowing and distance
between the corresponding user and antenna element, and $h^{[kl]}_{mji}\in \mathbb{C}$ is the small-scale fading coefficient. The received signal also includes additive white Gaussian noise. Since the
distance between a user and an array is much larger than the distance between the elements of an array, we assume that the
large-scale fading coefficients are independent of the antenna element index $m$. The small-scale fading coefficients, $h^{[kl]}_{mji}$, are assumed to be complex Gaussian zero-mean and unit-variance, and for any $(k,l,m,j,i) \neq (n,v,r,u,q)$ coefficients $h^{[kl]}_{mji}$ and $h^{[nv]}_{ruq}$ are independent.

%The small-scale fading coefficients, and the channel coefficients between $A_{ji}$ and $U_{kl}$ are represented by the following complex vectors, respectively:%\emph{small-scale fading vector}, as
%\begin{align*}
%\hbf^{[kl]}_{ji} &=\Big(h^{[kl]}_{1ji},h^{[kl]}_{2ji}, \ldots, h^{[kl]}_{Mji} \Big)^T \in \mathbb{C}^{M\times1},\\
%\gbf^{[kl]}_{ji} &= \Big(g^{[kl]}_{1ji},g^{[kl]}_{2ji}, \ldots, g^{[kl]}_{Mji} \Big)^T =
%\sqrt{\beta^{[kl]}_{ji}}\hbf^{[kl]}_{ji} \in \mathbb{C}^{M\times1}.
%\end{align*}

% and the channel coefficients between $A_{ji}$ and $U_{kl}$ form the \emph{channel vector}:
% \begin{align*}
% \gbf^{[kl]}_{ji}=\Big(g^{[kl]}_{1ji},g^{[kl]}_{2ji}, \ldots, g^{[kl]}_{Mji} \Big)^T= \sqrt{\beta^{[kl]}_{ji}}\hbf^{[kl]}_{ji} \in \mathbb{C}^{M\times1}.
% \end{align*}

%The small-scale fading coefficients, $h^{[kl]}_{mji}$, are assumed to be zero-mean and unit-variance, and for any $(k,l,m,j,i) \neq (n,v,r,u,q)$ coefficients $h^{[kl]}_{mji}$ and $h^{[nv]}_{ruq}$ are independent. It follows that, for any $(k,l,j,i)\neq(n,v,u,q)$, vectors $\hbf^{[kl]}_{ji}$ and $\hbf^{[nv]}_{uq}$ ($\gbf^{[kl]}_{ji}$ and $\gbf^{[nv]}_{uq}$) are independent, and $\hbf^{[kl]}_{ji} \sim \mathcal{CN}(\textbf{0},\textbf{I}_{M})$ ($\gbf^{[kl]}_{ji} \sim \mathcal{CN}(\textbf{0},\beta^{[kl]}_{ji}\textbf{I}_{M})$).
We will use $\gbf^{[kl]}_{ji} \in \mathbb{C}^{M\times1}$
to denote {\em channel vector} between $A_{ji}$ and $U_{kl}$.
We further assume that small-scale and large-scale fading coefficients are constant over small-scale and large-scale coherence blocks represented by $T$ and $T_\beta$ symbols, respectively. While the small-scale fading coefficients significantly change as soon as a user moves by a quarter of the wavelength, large-scale
fading coefficients are approximately constant in the radius of 10 wavelengths (see \cite{huang2011mimo} and references there). Thus, $T_\beta \approx 40~T$.
We also assume that small-scale channel coefficients are independent across different small-scale coherence blocks, and similarly large-scale channel coefficients.
%We remark that to address wide-band channels, orthogonal frequency-division multiplexing can be used in which our analysis is valid for each subcarrier.

\subsection{Time-Division Duplexing Protocol}\label{subsec: TDD protocol}
Uplink and downlink transmissions, require access to the channel vectors at the antenna arrays. Channel vectors are estimated by antenna arrays using uplink training transmissions in each small-scale coherence block $T$. Similar to \cite{marzetta2010noncooperative} and \cite{pcp1}, we assume that the same set of $K$ orthonormal training sequences (pilots) is reused in each cell, such that sequence $r^{[k]}\in \mathbb{C}^\tau$ is assigned to $U_{kl}$ in $C_l$, and $r^{[k]^\dagger}r^{[n]}=\delta_{kn}$ for any $k,n \in [K]$ and any $l\in[L]$. Note that since the number of orthogonal
$\tau$-tuples can not exceed $\tau$, we  have $K\leq\tau$\cite{pcp1}. Due to the independence of channel coefficients across different small-scale coherence blocks, training is repeated in each block $T$, hence $\tau \leq T$.
%For high-mobility users, the small-scale coherence block, $T$, is short, imposing a small training time $\tau$. %$\tau<T$,
 %we assign orthogonal training sequences to users within one cell, and since the number of orthogonal $\tau$-tuples can not exceed $\tau$, we have $K\leq\tau$, and reuse the same set in different cells.

The system operates based on the TDD protocol proposed in \cite{marzetta2010noncooperative},\cite{pcp1}. The first two steps of the protocol are carried out once for each large-scale coherence block, and the last five are repeated over small-scale coherence blocks.\\

\vspace{2pt}
\noindent{\bf Time-Division Duplexing Protocol}
%\begin{itemize}[noitemsep,nolistsep]%[noitemsep,topsep=0pt]

\noindent{\bf Step 1:} In the beginning of each large-scale coherence block, each base station estimates the large-scale fading coefficients between itself and all the users in the network.
%$A_{ji}$ estimates the large-scale fading coefficients $\beta^{[kl]}_{ji}$, $(k,l) \in [K]\times [L]$. \label{TDD:step1}

\noindent{\bf Step 2:} Each array transmits a measure of the large-scale fading coefficients estimated in Step $1$, to the users in its cell, which are later used for decoding the downlink signals in Step $7$. More specifically, $A_{ji}$, $i \in [3]$ transmits the {\em decoding coefficient} defined as
\begin{align}
\epsilon^{[kj]}_{ji} \triangleq
\frac{ \sqrt{ M  \rho_r \tau \rho^{[kj]}_{ji} }  G^{[kj]}_{ji}  \beta^{[kj]}_{ji}  }
{ \big( \sigma^2_r+ \rho_r \tau \sum_{l=1}^L  G^{[kl]}_{ji} \beta^{[kl]}_{ji} \big)^{1/2}}, \label{epsilon-dir}
\end{align}
to $U_{kj}$, $k\in[K]$, where $\sigma^2_r$ is the reverse link (uplink) noise power, $\rho_r$ is the reverse link transmit power from each user in $C_j$ to arrays $\{A_{ji}:i\in[3]\}$, and $\rho^{[kj]}_{ji}$ denotes the forward link (downlink) transmit power assigned by $A_{ji}$ to $U_{kj}$. Forward link power allocation strategies are discussed in Sec.~\ref{subsec: Downlink Power Allocation}.

\noindent{\bf Step 3:} All users synchronously transmit their uplink signals.

\noindent{\bf Step 4:} All users synchronously transmit their training sequences (pilots).

\noindent{\bf Step 5:} Each array estimates the channel vector between itself and the users located within its
cell using the training sequences, and processes the received uplink signals.

\noindent{\bf Step 6:} Arrays use conjugate beamforming (based on the estimated channel vectors and power allocation) in order to prepare the downlink signals $\{s^{[kj]}:k\in [K]\}$ for transmission, where $s^{[kj]}$ denotes the signal intended for $U_{kj}$. All arrays synchronously transmit the prepared signals. \label{TDD:step6}

\noindent{\bf Step 7:} User $U_{kj}$, $k,j\in[K]\times[L]$ decodes its received signal, denoted by $y^{[kj]}$, using the decoding coefficients received in Step $2$ as 
%$\hat{s}^{[kj]}=\frac{ y^{[kj]} }{\sum_{i=1}^3 \epsilon^{[kj]}_{ji} }$.
\begin{align}
\hat{s}^{[kj]}=\frac{ y^{[kj]} }{\sum_{i=1}^3 \epsilon^{[kj]}_{ji} }. \label{TDD:step7}
\end{align}

For the TDD protocol given above, we assume that each array $A_{ji}$ can accurately estimate and track all the large-scale fading coefficients, discussed in \cite{pcp1}, and it has the means to forward the decoding coefficients, $\epsilon^{[kj]}_{ji}$ to the users in $C_j$. As in \cite{pcp1}, we will not consider the resources needed for implementing these assumptions. 

\begin{Remark}
According to (\ref{epsilon-dir}), $\epsilon^{[kj]}_{ji}$ only depends on the large-scale fading coefficients the number of which, does not increase with the number of antennas as discussed in Sec. \ref{subsec:channel model}. Therefore, the amount of information exchange between each antenna array and its corresponding users does not depend on $M$, which makes the massive MIMO system scalable.
\end{Remark}

In the following we only analyze the downlink transmissions; the analysis of the uplink scenario follows similarly.
%is left for future research. 

% **************************************************************************************************
% Downlink System Performance
% **************************************************************************************************
%\vspace{-2mm}
\section{Downlink System Analysis} \label{sec: Downlink Analysis}
%\vspace{-2mm}
In this section, we analyze the downlink system performance by providing SINR expression. Theorem~\ref{Thm: downlink SINR} provides a lower bound on user downlink transmission rates. We assume that linear MMSE estimation is used to estimate the channel vectors in Step $5$ of the TDD protocol. Furthermore, as stated in step 2 of the TDD protocol, power assignments are represented explicitly.
 %Downlink signals $s^{[kj]}, (k,j)\in [K]\x[L]$ are assumed to be i.i.d. zero-mean unit-variance random variables.%( i.e.,
In our analysis, we assume that $\EE[s^{[kj]}] = 0$ and $\Var[s^{[kj]}] = 1$ for any $(k,j)\in [K]\x[L]$.
%\vspace{-5pt}
\subsection{Downlink System Performance}\label{subsec: Downlink Rate Performance}
\begin{theorem}\label{Thm: downlink SINR}
For the sectorized multi-cell massive MIMO system with directional antennas described in Sec. \ref{sec:system model}, the downlink transmission
rate to user $k\in [K]$ in cell $j\in [L]$, $R^{[kj]}$, is lower bounded by
% $U_{kj}, (k,j)\in [K]\times[L]$ is lower bounded by
\begin{align}
R^{[kj]} \geq \log_2(1+SINR^{[kj]}),
\end{align}
where,
\begin{align} \label{SINR}
%SINR^{[kj]} = & \frac{J_0^{[kj]}}{J_1^{[kj]}+ J_2^{[kj]} + 1},
SINR^{[kj]} = & \frac{P^{[kj]}}{I_1^{[kj]}+ I_2^{[kj]} + \sigma^2_f},
\end{align}
with,
%\begin{align}
%P^{[kj]} &= \Big| \sum_{i=1}^3 \sqrt{\rho_{ji}^{[kj]} G_{ji}^{[kj]}} \lambda_{ji}^{[kj]}\Big|^2 \label{j0},
%\end{align}
%\begin{align}
%I_1^{[kj]} &=  \sum_{\substack{l=1\\l\neq j}}^{L} \Big| \sum_{i=1}^3 \sqrt{\rho_{li}^{[kl]} G_{li}^{[kj]}} \lambda_{li}^{[kj]}\Big|^2 \label{j1},
%\end{align}
%
%\begin{align}
%I_2^{[kj]} &=  \sum\limits_{l=1}^{L}  \sum\limits_{i=1}^{3}  \rho_{li} G_{li}^{[kj]}  \beta_{li}^{[kj]} \label{j2},\\
%\lambda_{ji}^{[kl]} &= \bigg({\frac{ M \rho_r \tau G_{ji}^{[kl]} \beta_{ji}^{[kl]^2}}{\sigma^2_r+ \rho_r \tau \sum\limits_{v = 1}^{L} G_{ji}^{[kv]} \beta_{ji}^{[kv]}} }\bigg)^{1/2}, \label{lambda}
%\end{align}
\begin{align}
P^{[kj]} &= \Big| \sum_{i=1}^3 \sqrt{\rho_{ji}^{[kj]} G_{ji}^{[kj]}} \lambda_{ji}^{[kj]}\Big|^2 \label{j0}, \\
I_1^{[kj]} &=  \sum_{\substack{l=1\\l\neq j}}^{L} \Big| \sum_{i=1}^3 \sqrt{\rho_{li}^{[kl]} G_{li}^{[kj]}} \lambda_{li}^{[kj]}\Big|^2 \label{j1},\\
I_2^{[kj]} &=  \sum\limits_{l=1}^{L}  \sum\limits_{i=1}^{3}  \rho_{li} G_{li}^{[kj]}  \beta_{li}^{[kj]} \label{j2},\\
\lambda_{ji}^{[kl]} &= \bigg({\frac{ M \rho_r \tau G_{ji}^{[kl]} \beta_{ji}^{[kl]^2}}{\sigma^2_r+ \rho_r \tau \sum\limits_{v = 1}^{L} G_{ji}^{[kv]} \beta_{ji}^{[kv]}} }\bigg)^{1/2}, \label{lambda}
\end{align}
and $\rho_{li} \triangleq \sum_{k=1}^K \rho_{li}^{[kl]}$ is the forward link transmission power at array $i\in[3]$ in cell $j\in[L]$ and $\sigma^2_f$ denotes forward link noise power.
\end{theorem}

The sketch of the proof is provided in Appendix \ref{App: downlink SINR}.

%Due to the space limit, the proof is omitted.

%We refer the reader to Sec. IV of \cite{pcp1}, where a similar lower bound is provided for omnidirectional antennas at each base station.

%\begin{proof}
%Proof of Theorem \ref{Thm: downlink SINR} is given in Appendix \ref{App: downlink SINR}.
%\end{proof}

In Theorem~\ref{Thm: downlink SINR}, $P^{[kj]}$ is the desirable signal power received by $U_{kj}$, and $I_1^{[kj]}$, $I_2^{[kj]}$ correspond to two types of interference experienced by the user. More specifically, $I_1^{[kj]}$
is the interference created by pilot reuse in multiple cells, referred to as {\em pilot contamination}, and similar to
$P^{[kj]}$, it grows linearly with the number of base station antenna elements ($\lambda_{li}^{[kj]}$$\propto$$\sqrt{M}$).
The second interference $I_2^{[kj]}$, referred to as {\em undirected interference}, is created by nonorthogonality of channel vectors of different users, channel estimation error, and lack of user's knowledge of effective channel \cite{pcp1}. This type of interference does not grow with $M$, and hence has negligible contribution when $M$ is very large. Although increasing $M$ leads to higher SINR for all users, we remark that SINR converges to a bounded limit when $M$ goes to infinity.

%$I_2^{[kj]}$, referred to as {\em undirected interference}, is caused by channel estimation error, nonorthogonality of channel estimates of users using different training sequences, and the user effective channel uncertainty \cite{pcp1}. These types of interference do not grow with $M$, and hence have negligible contribution when $M$ is very large. Although increasing $M$ leads to higher SINR for all users, we remark that SINR converges to a bounded limit when $M$ goes to infinity.

%However, $I_2^{[kj]}$, {\BLUE {\em referred to as estimation contamination?}, corresponds to the interference imposed by the total
%network on user $U_{kj}$. This arises from the non-orthogonality of the channel estimates, which with the multi-user activity results in reception of undesired signals.} This type of interference is independent of the number of elements employed at the array, $M$.
%{\BLUE do we need to compare with pcp1?}

In the next section, we consider optimal and suboptimal strategies for forward link power allocation. In Sec~\ref{sec: Numerical Analysis} we evaluate the system performance and show that using optimized power allocation can lead to a significant performance improvement.

\subsection{Forward Link Power Allocation}\label{subsec: Downlink Power Allocation}
In Step $2$ and $6$ of the TDD protocol given in Sec.~\ref{subsec: TDD protocol}, arrays divide their forward link transmit power among the users they serve for downlink transmissions. In the following, we assume that $\rho_{li}=\sum_{k=1}^K \rho_{li}^{[kl]} \leq \rho_f/3$ for any $(l,i)\in[L]\x[3]$, where $\rho_f$ is the base station maximum forward link power, and discuss three different strategies with different communication and computation complexities, and compare their performance in Sec.~\ref{sec: Numerical Analysis}.

\subsubsection{Uniform Power Allocation (UPA)} \label{subsubsec: Uniform Power Allocation}
In this suboptimal strategy, which requires no cooperation across the network, each array transmits at full power and divides its forward link transmit power uniformly across the users in its cell such that each gets a portion equal to $\rho_{ji}^{[kj]}=\frac{\rho_f}{3K}$, $\forall(j,i,k)\in[L]\times[3]\times[K]$.

%{\BLUE We remark that, with UPA the SINR expression can be explicitly written in terms of $M$ and $K$, and the $M\gg K$ assumption guarantees that, in the absence of pilot contamination, user downlink transmission rate grows without bound as the number of base station antenna elements grows infinitely.(do we need to discuss the absence of pilot contamination here?)}
%{\RED OR}\\
%{\BLUE With UPA, $P^{[kj]}$ and $I_1^{[kj]}$ are proportional to $M/K$ ($\lambda_{li}^{[kj]}\propto\sqrt{M}$ and $\rho_{li}^{[kl]}\propto1/K$) and $I_2^{[kj]}$ is constant ($\rho_{li}=\rho_f/3$) for every $k \in [K]$ and $j \in [L]$. Therefore, SINR expression introduced in (\ref{SINR}) is an increasing function of $M/K$ for every user and it converges to its limit when $M/K \gg 1$ or $M \gg K$. Therefore, given UPA, assumption $M \gg K$ guarantees a near optimal performance for the system.}

%\subsubsection{Locally Optimized Power Allocation (Local PA)} \label{subsubsec: Local Power Allocation}
%{\BLUE keep?}
%either in a decentralized fashion or by a central unit based on system knowledge,

\subsubsection{Optimal Centralized Power Allocation (CPA)} \label{subsubsec: CPA}
The powers allocated to each user can be globally optimized in order to maximize the worst downlink SINR (equivalently rate) among all users in the network. A central entity formulates and solves a constrained max-min optimization problem based on the SINR expression given in Theorem \ref{Thm: downlink SINR} as follows, ensuring to satisfy each array's maximum forward link transmit power. %Due to space limitations we do not give the full details of the problem formulation and only provide the following relaxed optimization problem (please see \cite{pcp1}).
%\begin{subequations}
\begin{align}\label{max-min-sinr}
&\max_{ \{\rho_{li}^{nl}\}} \; \min_{k,j} \frac{P^{[kj]}}{I_1^{[kj]} + I_2^{[kj]} + \sigma^2_f}    \qquad\qquad\quad \\
&\text{subject to:} \notag \\
& \qquad \sum\limits_{n=1}^K \rho_{li}^{[nl]} \leq \frac{\rho_f}{3}, \;\;\; \forall (l,i)\in [L]\times[3] \notag\\
& \qquad \rho_{li}^{[nl]} \geq 0 ,\;\;\; \forall (l,i,n)\in [L]\times[3]\times [K] \notag
\end{align}
%\end{subequations}
with $P^{[kj]}$, $I_1^{[kj]}$, and $I_2^{[kj]}$ given in (\ref{j0})-(\ref{j2}). By introducing slack variables $X_{kj}$ and $Y_{kj}$, (\ref{max-min-sinr}) is equivalent to:
 \begin{align}
&\max_{ \{\psi_{li}^{nl},X_{nl},Y_{nl}\} }
\min_{k,j} \frac{\Big| \sum_{i=1}^3 \psi_{ji}^{[kj]}\sqrt{ G_{ji}^{[kj]}} \lambda_{ji}^{[kj]}\Big|^2}{ X_{kj}^2 +  Y_{kj}^2 + \sigma^2_f} \label{max-min-sinr-relaxed}\\
&\text{subject to: } \notag\\
&\sum_{\substack{l=1\\l\neq j}}^{L} \Big| \sum_{i=1}^3 \psi_{li}^{[kl]}\sqrt{ G_{li}^{[kj]}} \lambda_{li}^{[kj]}\Big|^2 \leq X_{kj}^2,\;\;\; \forall(j,k)\in[L]\times[K], \notag\\
&\sum\limits_{l=1}^{L} \sum\limits_{n=1}^{K} \sum\limits_{i=1}^{3} (\psi_{li}^{[nl]})^2 G_{li}^{[kj]}  \beta_{li}^{[kj]} \leq Y_{kj}^2,\;\;\; \forall(j,k)\in[L]\times[K], \notag\\
&\sum\limits_{k=1}^K (\psi_{li}^{[kl]})^2 \leq \frac{\rho_f}{3}  ,\;\;\; \forall (l,i)\in [L]\times[3],\notag\\
&\psi_{li}^{[kl]} \geq 0, \;\;\; \forall(l,i,k)\in[L]\times[3]\times[K]\notag,
\end{align}
where, $\psi_{ji}^{[kj]}=\sqrt{\rho^{[kj]}_{ji}}$. The equivalence between (\ref{max-min-sinr}) and (\ref{max-min-sinr-relaxed}) follows from the fact that first two constraints in (\ref{max-min-sinr-relaxed}) hold with equality at the optimum.

\begin{proposition}
\label{Thm:quasiconcavity}
Power optimization problem (\ref{max-min-sinr-relaxed}) is quasi-concave.
\end{proposition}

The proof of proposition \ref{Thm:quasiconcavity} is provided in Appendix \ref{App:quasiconcavity}.

Due to quasi-concavity of problem (\ref{max-min-sinr-relaxed}), the solution can be obtained using the bisection method and a series of feasibility checking convex problems provided in Algorithm \ref{bisection}.

%It can be shown that (\ref{max-min-sinr-relaxed}) is a quasiconcave optimization problem, and the optimal power allocation can be computed using the bisection method and a series of feasibility checking convex problems provided in Algorithm \ref{bisection}.

%Due to space limitations we do not provide the full details of the equivalent optimization problem formed by introducing slack variables as done in \cite{ngo2016cell}, and only provide Algorithm \ref{bisection} from which the optimal power allocation is derived as $\rho^{[nl]^*}_{li} = \psi^{[nl]^{*2}}_{li}$.
\begin{algorithm}[h]
\caption{Centralized Power Allocation}
 \label{bisection}
\begin{algorithmic}[1]
\STATE %\emph{Initialization}:
Choose a tolerance threshold $ \delta > 0$, and initialize $\gamma_{min}$ and $\gamma_{max}$ to define a range of relevant values of the objective function in (\ref{max-min-sinr-relaxed}).
%, where $t_{min}$ and $t_{max}$ define a range of relevant values of the objective function in \eqref{max-min-sinr-relaxed}.
\WHILE {$\gamma_{max}-\gamma_{min} > \delta$}
    \STATE %\emph{Feasibility checking}:
    Set $\gamma := \frac{\gamma_{min}+\gamma_{max}}{2}$ and solve the following convex feasibility checking problem for $V_{kj} \triangleq [X_{kj},Y_{kj},1]$:
    \begin{align}
    & \scalebox{0.92}{$  ||V_{kj}|| \leq \frac{1}{\sqrt{\gamma}} \sum_{i=1}^3 \psi_{ji}^{[kj]}\sqrt{ G_{ji}^{[kj]}} \lambda_{ji}^{[kj]} ,\, \forall(j,k)\in[L]\times[K],  $} \notag \\
    %\label{feas-check}\\
    & \scalebox{0.92}{$ \sum\limits_{\substack{l=1\\l\neq j}}^{L} \Big| \sum\limits_{i=1}^3 \psi_{li}^{[kl]}\sqrt{ G_{li}^{[kj]}} \lambda_{li}^{[kj]}\Big|^2 \leq X^2_{kj},\, \forall(j,k)\in[L]\times[K],   $}
    \notag\\
    & \scalebox{0.92}{$\sum\limits_{l=1}^{L} \sum\limits_{n=1}^{K} \sum\limits_{i=1}^{3} (\psi_{li}^{[nl]})^2 G_{li}^{[kj]}  \beta_{li}^{[kj]} \leq Y^2_{kj},\; \forall(j,k)\in[L]\times[K],  $}
    \notag\\
    & \scalebox{0.92}{$ \sum\limits_{k=1}^K (\psi_{li}^{[kl]})^2 \leq \frac{\rho_f}{3} , \;\forall (l,i)\in [L]\times[3],    $}
    \notag\\
    & \scalebox{0.92}{$ \psi_{li}^{[kl]} \geq 0, \;\forall(l,i,k)\in[L]\times[3]\times[K],     $}
    \notag
    \end{align}
    \IF {above problem is feasible}
        \STATE $\gamma_{min}:=\gamma$
    \ELSE
        \STATE $\gamma_{max}:=\gamma$
    \ENDIF

\ENDWHILE

\end{algorithmic}
\end{algorithm}

This power allocation strategy requires a central entity with access to the large-scale fading coefficients of the entire network, and has a much higher complexity compared to the UPA scenario. In this scheme, in each large-scale coherence block $T_\beta$, the base stations send their estimated large-scale fading coefficients to a central entity. The central entity solves the optimization problem using Algorithm \ref{bisection}, and sends the results back to the base stations.

\subsubsection{Decentralized Power Allocation (DPA)} \label{subsubsec: DPA}
Computational and communication complexities of CPA can be significantly reduced using an optimization based on local information. Each antenna array, say $A_{ji}$, considers itself and the antenna arrays in a ring of cells around $C_j$ such that if this would be the entire network. $A_{ji}$ collects all large-scale fading channel coefficients between all antennas arrays and all users in this network and solves the optimization problem (\ref{max-min-sinr}), formulated for this network. Next, $A_{ji}$ uses the found downlink powers $\rho_{li}^{[kl]},~\forall k\in [K]$, and discard the powers found for other antenna arrays in the ring.

%Each base station, say base station $b$ located at corner $i$ of cell $l$, considers itself and the base stations in a ring of cells around cell $l$ such that if this would be the entire network. Base station $b$ collects all large-scale fading channel coefficients between all base stations and all users in this network and solves the optimization problem (\ref{max-min-sinr}), formulated for this network. Next, base station $b$ uses the found downlink powers $\rho_{li}^{[kl]},~\forall k\in [K]$, and discard other found powers.

\section{Simulations and Discussions}\label{sec: Numerical Analysis}
%In this section, we numerically discuss how effective sectoring is in mitigating interference in massive MIMO
%systems. In our simulations we consider the following sectorized settings: {\em Directional
%Arrays with UPA (Dir-UPA)} and {\em Directional Arrays with CPA (Dir-CPA)}, and compare them with their omnidirectional counterparts: {\em Omnidirectional Base Stations with UPA (Omni-UPA)} and {\em Omnidirectional Base Stations with CPA(Omni-CPA)}. The system model in settings Dir-UPA and Dir-CPA is the one introduced in Sec~\ref{sec:system model}, while the other two settings are modeled based on the setting in \cite{pcp1}, where one base station, equipped with $M_{B}$ omnidirectional antenna elements, is placed at the center of the cell, and each have a forward link power budget of $\rho_f$. For each setting, the forward link powers are allocated according to the their respective strategies defined in Sec.~\ref{subsec: Downlink Power Allocation}.

In this section, we evaluate how effective sectoring is in mitigating the interference in massive MIMO
systems. In our simulations we consider the following sectorized settings: {\em Directional
Arrays with UPA (Dir-UPA)}, {\em Directional Arrays with CPA (Dir-CPA)}, {\em Directional Arrays with DPA (Dir-DPA)}, and compare them with their omnidirectional counterparts: {\em Omnidirectional Base Stations with UPA (Omni-UPA)}, {\em Omnidirectional
Base Stations with CPA (Omni-CPA)}, and {\em Omnidirectional Base Stations with DPA (Omni-DPA)}. The system model in settings Dir-UPA, Dir-CPA and Dir-DPA is the one introduced in Sec~\ref{sec:system model}, while the other three settings are modeled based on \cite{pcp1}, where one base station with $M_{B}$ omnidirectional antenna elements, is placed at the center of the cell, and has a forward link power budget of $\rho_f$. For each setting, the forward link powers are allocated according to their respective strategies defined in Sec.~\ref{subsec: Downlink Power Allocation}.

   \begin{figure*}
 \centering \includegraphics[width=\textwidth]{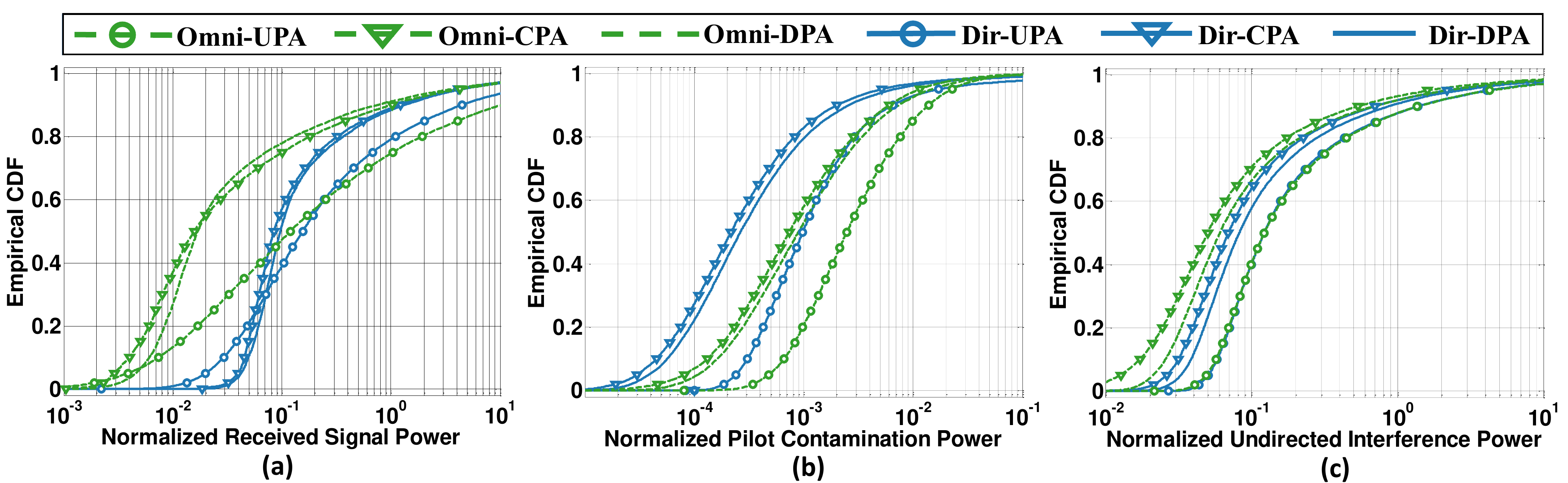}
 \caption{CDF of (a) normalized received signal power, (b) normalized pilot contamination power, and (c) normalized undirected interference power, for $M_B=10^2$.}
  \label{fig:simulations1}
 \end{figure*}

We consider a network composed of $L = 19$ cells (two rings of cells around a central cell), each with a radius of $R=1~km$,
and $K = 9$ users distributed uniformly across each cell except for a disk with radius $r = 60~m$ around the base stations. In order to avoid the cell edge effect, cells are wrapped into a torus as in \cite{wrap,pcp1}. The
large-scale fading coefficients are modeled based on the `COST-231' model at central frequency $f_c=1900~MHz$ as $10\log_{10} \big( \beta^{[kj]}_{li}
\big)=~-140-35.2\log_{10}(d^{[kj]}_{li})+\Psi$, where $d^{[kj]}_{li}$ denotes the distance (in $km$) between $U_{kj}$ and
$A_{li}$, and $\Psi$ denotes the shadow fading coefficient. We assume that $\Psi \sim \mathcal{N}\big(0,8\big)$, thermal noise power is $-101~dBm$, and the noise figure at each base station and each user is $9~dB$, hence $\sigma^2_f=\sigma^2_r=-92~dBm$. The antenna main-lobe and back-lobe power gains are $G_Q=2.98$ and $G_q=0.01$, respectively, the reverse link transmit
power is $\rho_r = 23~dBm$, and the maximum forward link transmit power of each base station is set at $\rho_f = 30~dBm$.
%which is assumed to be equally divided among the three arrays in sectorized settings.

Figs.~\ref{fig:simulations1}(a)-\ref{fig:simulations1}(c) display the CDF of the normalized received signal power where normalization with respect to forward link noise power, i.e. $P^{[kj]}/\sigma^2_f$, and normalized version of two types of interference powers affecting the users in a network, i.e. $I_1^{[kj]}/\sigma^2_f$ and $I_2^{[kj]}/\sigma^2_f$. We only
provide the simulations for $M_B = 10^2$, since, as seen in Theorem~\ref{Thm: downlink SINR} in Sec.~\ref{subsec: Downlink Rate
Performance}, received signal power and pilot contamination power are linearly proportional to $M$, while undirected interference power is independent of $M$. When comparing the Dir-UPA and Omni-UPA settings, we observe that sectoring affects each of these components as follows:
\subsubsection{Received signal power $P^{[kj]}$}
%\noindent\textbf{Received signal power $P^{[kj]}$}:
With sectoring, received signal power is higher for most of the users. In Dir-UPA, each user communicates with three arrays, each of which has $M = M_B/3$ elements and a forward link transmit power of $\rho_f/3$. Even though the per-element forward link transmit power is equal to that in Omni-UPA, i.e. $\rho_f/M_B$, users benefit from
the directionality of the antenna arrays. In this case, the signals transmitted from each array are emitted with the
main-lobe directionality gain ($G_Q \approx 3$), compared to the unity directionality gain of an omnidirectional base station.

Another reason for the increase in the received signal power is reduction of the pilot contamination effect (see the next subsection). The pilot contamination has two malicious effects. First, a base station creates directed interference to users located in other cells. Second, since the base station deviates part of its transmit power to other users, it effectively reduces the transmit power for users located in its cell. With sectoring the pilot contamination effect is getting smaller (see the next subsection), and therefore the signal power for legitimate users increases.

The net gain translates into an increase in the received signal power of $60\%$ of the users.

\subsubsection{Pilot contamination $I_1^{[kj]}$}
\begin{figure}
\centering
 \includegraphics[width=2.2in]{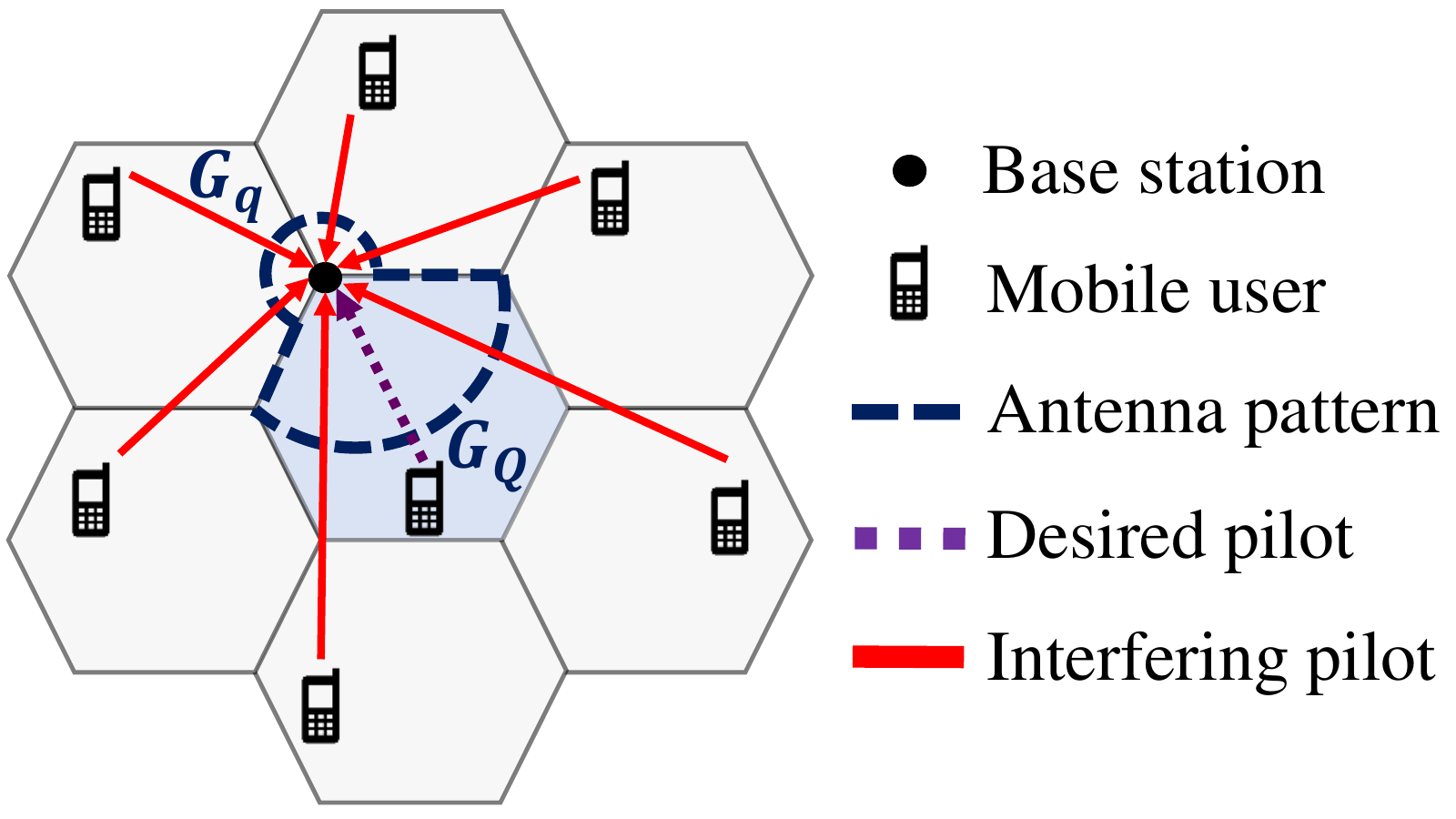}
\caption{Pilot contamination in sectorized networks.}\label{fig: pilot contamination}
\vspace{-10pt}
\end{figure}

%\noindent\textbf{Pilot contamination $I_1^{[kj]}$}:
Sectoring reduces the effect of pilot contamination. This is due to the fact that with the
directionality in Dir-UPA, arrays are able to derive better channel estimates from the received pilots, and further mitigate the pilot contamination. In Omni-UPA, each array receives the pilots transmitted from all cells and in all directions. However, directional arrays receive these signals with different directionality gains from the users in
different cells, i.e., one-third of the signals (those in the main lobe coverage of the arrays) are amplified with $G_Q \approx 3$, while the remaining two-thirds (in the back lobe coverage of the arrays) are attenuated with $G_q \approx 0$ as illustrated in Fig.
\ref{fig: pilot contamination}. In this case, the effective channel estimation SINR is approximately $3$ times larger compared to
the omnidirectional setting, which in turn, as depicted in Fig.~\ref{fig: pilot contamination}, reduces the interference.
% $\frac{1}{3}^{\text{rd}}$ ?

\subsubsection{Undirected interference $I_2^{[kj]}$}
%\noindent\textbf{Undirected interference $I_2^{[kj]}$}:
Sectoring does not affect undirected interference power. In both Dir-UPA and Omni-UPA settings, the multi-user activity in the overall network contributes to the undirected interference power, which arises due to the nonorthogonality of the channel vectors and other parameters mentioned in Sec.~\ref{subsec: Downlink Rate Performance}. More specifically, in the sectorized scenario, all of the $M_BL$ antennas create interference, among which, each user receives the signals emitted from one-third
amplified by a factor of $G_Q \approx 3$, and signals from the remaining antennas are attenuated by $G_q \approx 0$. Therefore, in sectorized scenario there are $M_BL/3$ effective antenna elements in the network contributing to $I_2^{[kj]}$ by transmitting their downlink signals with power $\rho_f/M_B$ amplified by $G_Q \approx 3$, creating the same amount of interference compared to the omnidirectional setting, where there are $M_BL$ antenna elements contributing to $I_2^{[kj]}$ by transmitting their downlink signals with power $\rho_f/M_B$.
%creating the same amount of interference compared to the omnidirectional setting.

We observe that for both directional and omnidirectional antenna settings, with CPA and DPA, the received signal power is higher for low-SINR users, and interference power is less for all users compared with their UPA counterparts. We remark that the difference in the performance of DPA and CPA is small for both settings.

%  \begin{figure*}
% \centering \includegraphics[scale=0.9]{Plot22.pdf}
% \caption{CDF of achievable downlink rates (a) $M_B=10^2$, (b) $M_B=10^4$ and (c) $M_B=10^6$.}
%  \label{fig:simulations2}
% \end{figure*}

 \begin{figure*}[t]
 \centering \includegraphics[width=\textwidth]{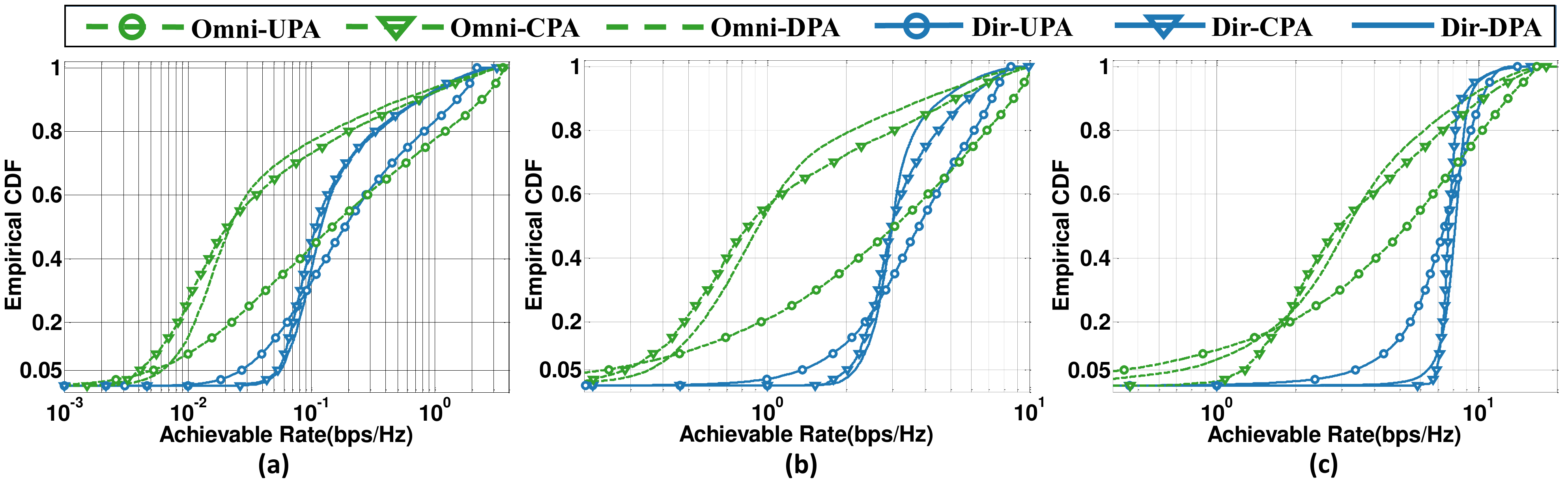}
 \caption{CDF of achievable downlink rates (a) $M_B=10^2$, (b) $M_B=10^4$ and (c) $M_B=10^6$.}
  \label{fig:simulations2}
 \end{figure*}

We provide CDFs of the downlink achievable rates for sectoring, given in Theorem \ref{Thm: downlink SINR}, and compare them for different settings in Figs.~\ref{fig:simulations2}(a)-(c), for $M_B=10^2$, $M_B=10^4$, and $M_B =10^6$, respectively. For comparison we use the {\em$0.95$-likely rate} per user criterion, defined as the rate achieved by $95\%$ of the users, as in \cite{marzetta2010noncooperative,pcp1,pcp2}.

For small values of $M_B$, the total interference imposed on $U_{kj}$ is dominated by undirected interference, which is similar for settings with and without sectoring. Therefore, directional arrays increase user SINR due to the increase in their received signal powers. For example with $M_B=10^2$, comparing the performance of Dir-UPA with Omni-UPA, given in Fig.~\ref{fig:simulations2}(a), we observe that sectoring is able to increase the \emph{$0.95$-likely} rate by a factor of $5.20$. We remark that as argued in Fig.~\ref{fig:simulations2}(a) in Dir-UPA, the achievable rate of around $60\%$ of the users with lower SINR has been improved with a sacrifice from the rate of user with higher SINR. %even though the achievable rate of $38\%$ of the high-SINR users is reduced, a larger percentage of the low-SINR users achieve a large increase in their rates.
For intermediate $M_B$, the two types of interference are comparable, and therefore, in addition to the increase in received signal power, directional arrays are able to alleviate the effect of the total interference. As illustrated in Fig.~\ref{fig:simulations2}(b), for $M_B=10^4$ the \emph{$0.95$-likely} rate has an improvement with a factor of $5.47$ with the Dir-UPA compared to Omni-UPA, and the rate of $66\%$ of the users is increased.
In the regime of very large $M_B$, pilot contamination is dominant, and therefore, as $M_B\rightarrow \infty$, SINR converges to a finite value. For $M_B=10^6$, given in Fig.~\ref{fig:simulations2}(c), the \emph{$0.95$-likely} rate in Dir-UPA is $7.65\times$ higher compared to Omni-UPA, with an improvement in achievable rate for $72\%$ of the users.

A comparison among Dir-UPA, Dir-CPA, and Dir-DPA for different $M_B$ in Fig. \ref{fig:simulations2} reveals that optimized power allocation schemes can improve \emph{$0.95$-likely} rate by a factor between $1.48$ and $2.04$.

We also would like to note that empirical CDF of achievable rate with decentralized power allocation (Dir-DPA) is only marginally different from the CDF of the optimal centralized power allocation (Dir-CPA), while using Dir-DPA allows us to reduce the required computation and communication overheads significantly.

\section{Conclusions}\label{sec: Conclusion}
In this paper, we have studied the benefits of using directional antennas at the base station in a massive MIMO system. We have derived a lower bound on user downlink achievable rates, and have discussed centralized and decentralized power allocation strategies by formulating power optimization problems which differ in terms of performance and complexity. We have compared the performance of different massive MIMO settings with and without sectoring, and for different power allocation methods in terms of received signal power, pilot contamination, undirected interference and their achievable rate. The numerical results have revealed that while sectoring does not affect the undirected interference, it can alleviate the effect of pilot contamination and increase received signal power. Finally, we have discussed how sectoring and the use of directional antennas leads to higher {\em$0.95$-likely} rate as a measure of reliability in the system. We have observed that by increasing the number of antennas at each base station, the improvement due to sectoring increases, due to the reduction of pilot contamination which is proportional to the number of antennas. Based on our simulation results, power optimization is an effective way to increase the {\em$0.95$-likely} rate further.

% We observe that sectoring improves user DL rates significantly. In fact, the improvement in the DL rate and reliability due to sectoring increases as $m$ increases. Furthermore, we observe that power optimization is an effective way (in both case with and without sectoring) to increase the reliability of the system.

% do not consider precoding, but improve the system performance by formulating and solving a power optimization problem. The simulation results in Sec. \ref{sec: Numerical Analysis} show how sectoring can increase user rates as an alternative method to reduce the interference.

% comment how power optimization is good or bad
% at the expense of a small rate sacrifice by the
% **************************************************************************************************
% Appendix
% **************************************************************************************************
%\vspace{-0.65cm}
\begin{appendices}
\section{Proof of Theorem \ref{Thm: downlink SINR}}\label{App: downlink SINR}
Due to space limit, we only provide a sketch of the proof here. 
%The derivation of the SINR expression follows that given in \cite{pcp1}, modified to account for the directivity of arrays, and the fact that users communicate with three base stations.
As described in the TDD protocol given in Sec.~\ref{subsec: TDD protocol}, once the arrays have estimated the large-scale fading coefficients (step 1) and transmitted the decoding coefficients to their users (step 2), all users synchronously transmit their uplink signals and training sequences, respectively, in steps 3 and 4. Then, in step $5$, each array estimates its channel vector using an MMSE estimate. More specifically, $A_{ji}$ estimates the channel vector $\gbf_{ji}^{[kl]}$ as:
\begin{align}
\hat\gbf_{ji}^{[kl]} %&=  \theta_{ji}^{[kl]} \big( \Ybf_{ji} \x  \rbf^{[k]} \big) \notag  \\
&=\theta_{ji}^{[kl]} \sqrt{\rho_r \tau} \sum_{v=1}^L \sqrt{G_{ji}^{[kv]}} \gbf_{ji}^{[kv]} +\hat{\wbf}_{ji}^{[kl]},
\end{align}
where,
\begin{align}
\theta_{ji}^{[kl]} =\frac{\sqrt{\rho_r \tau G_{ji}^{[kl]}} \beta_{ji}^{[kl]}}{\sigma_r^2+ \rho_r \tau \sum\limits_{v = 1}^{L} G_{ji}^{[kv]} \beta_{ji}^{[kv]}},
\end{align}
and, $\hat{\wbf}_{ji}^{[kl]}\sim \mathcal{CN} ( 0, \theta_{ji}^{[kl]^2} \Ibf_M)$, where $\Ibf_M$ is $M\times M$ identity matrix.
%\begin{align}
%\hat{\wbf}_{ji}^{[kl]}\sim CN ( 0, \theta_{ji}^{[kl]^2} \Ibf_M).
%\end{align}
We assume that $\gbf_{ji}^{[kl]} = \hat\gbf_{ji}^{[kl]} + \tilde\gbf_{ji}^{[kl]}$
where $\tilde\gbf_{ji}^{[kl]}$ denotes the MMSE estimation error.
It can be shown that %$\hat\gbf_{ji}^{[kl]}$ and $\tilde\gbf_{ji}^{[kl]}$ have the following distributions,
\begin{align}
&\hat\gbf_{ji}^{[kl]} \sim CN \Big( 0,  \frac{1}{M}\lambda_{ji}^{[kl]^2} \Ibf_M\Big),  \label{ghat-dir}\\
&\tilde\gbf_{ji}^{[kl]} \sim CN \Big( 0, \Big(\beta_{ji}^{[kl]}- \frac{\lambda_{ji}^{[kl]^2}}{M} \Big)\Ibf_M \Big), \label{gtild-dir}
\end{align}
where, $\lambda_{ji}^{[kl]}$ is given in (\ref{lambda}).
%\begin{align}
%{\RED\lambda_{ji}^{[kl]^2} = \EE[||\hat\gbf_{ji}^{[kl]}||^{2}] = {\frac{M \rho_r \tau G_{ji}^{[kl]} \beta_{ji}^{[kl]^2}}{1+ \rho_r \tau \sum\limits_{v = 1}^{L} G_{ji}^{[kv]} \beta_{ji}^{[kv]}} }. \label{lambda-dir}\text{{\BLUE we introduced this in the theorem}}}
%\end{align}
In step $6$, array $(j,i)\in [L]\times [3]$ uses conjugate beamforming based on its channel estimates to transmit the downlink
signals $\{s^{[kj]}: k\in[K]\}$ to its users, as
\begin{align}
\xbf_{ji} =\sum\limits_{k=1}^{K}   \frac{\sqrt{\rho_{ji}^{[kj]}}}{\lambda_{ji}^{[kj]}} \hat\gbf_{ji}^{[kj]^{\dagger}} s^{[kj]},
\end{align}
where $\rho_{ji}^{[kj]}$ denotes the power allocated to $U_{kj}$ by $A_{ji}$. User $U_{kj}$ receives the following downlink signal:
%\begin{align}
%y^{[kj]}
%&=  \sum\limits_{i=1}^{3} \frac{\sqrt{\rho_{ji}^{[kj]} G_{ji}^{[kj]}}}{\lambda_{ji}^{[kj]}} \EE\Big[  \hat\gbf_{ji}^{[kj]^{\dagger}}  \hat\gbf_{ji}^{[kj]} \Big] s^{[kj]}  \label{eq:T0} \\
%& \; +  \sum\limits_{i=1}^{3} \frac{\sqrt{\rho_{ji}^{[kj]} G_{ji}^{[kj]}}}{\lambda_{ji}^{[kj]}} \Big( \hat\gbf_{ji}^{[kj]^{\dagger}}  \hat\gbf_{ji}^{[kj]} - \EE\Big[  \hat\gbf_{ji}^{[kj]^{\dagger}}  \hat\gbf_{ji}^{[kj]} \Big]    \Big)   s^{[kj]}  \label{eq:T1} \\
%& \; +  \sum\limits_{\substack{l=1\\ l\neq j}}^{L} \sum\limits_{i=1}^{3}
% \frac{\sqrt{\rho_{li}^{[kl]}G_{li}^{[kj]}}}{\lambda_{li}^{[kl]}} \hat\gbf_{li}^{[kl]^{\dagger}}  \hat\gbf_{li}^{[kj]}  s^{[kl]}  \label{eq:T2} \\
%& \; +  \sum\limits_{l=1}^{L} \sum\limits_{i=1}^{3} \sum\limits_{\substack{n=1\\ n\neq k}}^{K}
% \frac{\sqrt{\rho_{li}^{[nl]}G_{li}^{[kj]}}}{\lambda_{li}^{[nl]}} \hat\gbf_{li}^{[nl]^{\dagger}}  \hat\gbf_{li}^{[kj]}  s^{[nl]} \label{eq:T3} \\
%& \; + \sum\limits_{l=1}^{L}\sum\limits_{i=1}^{3} \sqrt{G_{li}^{[kj]}} \xbf_{li} \tilde\gbf_{li}^{[kj]}  + w^{[kj]} \label{eq:T4}
%\end{align}

\begin{align}
y^{[kj]}
&= \sum\limits_{l=1}^{L}\sum\limits_{i=1}^{3} \sqrt{G_{li}^{[kj]}} \xbf_{li} \gbf_{li}^{[kj]} + w^{[kj]} \notag\\
&= \underbrace{\sum\limits_{i=1}^{3} \frac{\sqrt{\rho_{ji}^{[kj]} G_{ji}^{[kj]}}}{\lambda_{ji}^{[kj]}} \EE\Big[  \hat\gbf_{ji}^{[kj]^{\dagger}}  \hat\gbf_{ji}^{[kj]} \Big] s^{[kj]}}_{T_0} \notag\\ %\label{eq:T0} \\
&+  \underbrace{\sum\limits_{i=1}^{3} \frac{\sqrt{\rho_{ji}^{[kj]} G_{ji}^{[kj]}}}{\lambda_{ji}^{[kj]}} \Big( \hat\gbf_{ji}^{[kj]^{\dagger}}  \hat\gbf_{ji}^{[kj]} - \EE\Big[  \hat\gbf_{ji}^{[kj]^{\dagger}}  \hat\gbf_{ji}^{[kj]} \Big]    \Big)   s^{[kj]}}_{T_1} \notag\\  
%\label{eq:T1} \\
&+ \underbrace{\sum\limits_{\substack{l=1\\ l\neq j}}^{L} \sum\limits_{i=1}^{3}
 \frac{\sqrt{\rho_{li}^{[kl]}G_{li}^{[kj]}}}{\lambda_{li}^{[kl]}} \hat\gbf_{li}^{[kl]^{\dagger}}  \hat\gbf_{li}^{[kj]}  s^{[kl]}}_{T_2} \notag \\  %\label{eq:T2} \\
&+ \underbrace{\sum\limits_{l=1}^{L} \sum\limits_{i=1}^{3} \sum\limits_{\substack{n=1\\ n\neq k}}^{K}
 \frac{\sqrt{\rho_{li}^{[nl]}G_{li}^{[kj]}}}{\lambda_{li}^{[nl]}} \hat\gbf_{li}^{[nl]^{\dagger}}  \hat\gbf_{li}^{[kj]}  s^{[nl]}}_{T_3} \notag \\ %\label{eq:T3} \\
&+ \underbrace{\sum\limits_{l=1}^{L}\sum\limits_{i=1}^{3} \sqrt{G_{li}^{[kj]}} \xbf_{li} \tilde\gbf_{li}^{[kj]}  + w^{[kj]}}_{T_4}, \notag % \label{eq:T4}
\end{align}
%We refer to the terms in \eqref{eq:T0}, \eqref{eq:T1}, \eqref{eq:T2}, \eqref{eq:T3}, and \eqref{eq:T4}
%as $T_0$, $T_1$, $T_2$, $T_3$, and $T_4$, respectively. %Using (\ref{ghat-dir}) and (\ref{lambda-dir}), it follows that:
% \begin{align}
% T_0=s_D^{[kj]}\sum_{i=1}^3 \epsilon^{[kj]}_{ji},
% \end{align}
% with $\epsilon^{[kj]}_{ji}$ defined in (\ref{epsilon-dir}).
where, $w^{[kj]} \sim \mathcal{CN} \big(0, 1\big)$ denotes the noise. $T_0$ corresponds to the part of the received signal that $U_{kj}$ can decode, while $T_1,\dots, T_4$ contribute to the interference and noise. More specifically, using (\ref{lambda}) and (\ref{ghat-dir}), it can be shown that $T_0=s^{[kj]}\sum_i \epsilon^{[kj]}_{ji}$, and given $\{\epsilon^{[kj]}_{ji}:i \in [3]\}$, user $U_{kj}$ can decode $T_0$ using (\ref{TDD:step7}) to find $\hat{s}^{[kj]}$ in step 7 of TDD protocol. 
%Furthermore, we would like to note that $\epsilon^{[kj]}_{ji}$ only depends on the large-scale fading coefficients the number of which does not increase with the number of antennas, $M$. Therefore, the amount of information exchange between each antenna array and its corresponding users does not depend on $M$, which makes the system scalable.  
Furthermore, it can be shown that any two of the terms $T_0,\dots, T_4$ are uncorrelated. According to Theorem 1 of \cite{hassibi}, the worst case of uncorrelated additive noise is independent Gaussian
noise with the same variance. Hence, the worst-case downlink SINR of the of $U_{kj}$, denoted by $SINR^{[kj]}$, is
\begin{align}
SINR^{[kj]}  = \frac{\Var[T_0]}{\Var[T_1]+\Var[T_2]+\Var[T_3]+\Var[T_4]}. \label{SINR}
\end{align}
%According to Theorem 1 in \cite{hassibi}, the worst-case uncorrelated additive noise is independent Gaussian noise with the same variance.
Therefore, $U_{kj}$'s downlink rate, $R^{[kj]}$, is lower bounded by
\begin{align}
R^{[kj]}&= I (y^{[kj]};s^{[kj]} \mid \epsilon^{[kj]}_{j1},\epsilon^{[kj]}_{j2},\epsilon^{[kj]}_{j3}  )
 \geq \log_2 ( 1+SINR^{[kj]} ). \notag%\label{DL-rate-lowerbound}
\end{align}

It is straightforward to calculate variance of the terms $T_0,\dots, T_4$ based on (\ref{lambda}), (\ref{ghat-dir}), (\ref{gtild-dir}), and the channel statistics. Substituting these terms in (\ref{SINR}) will conclude the proof.

\section{Proof of Proposition \ref{Thm:quasiconcavity}}\label{App:quasiconcavity}

%The proof is similar to proof of Proposition 1 in \cite{ngo2017cell}. 
The constraints of problem (\ref{max-min-sinr-relaxed}) are convex. To prove the quasi-concavity, it suffices to show that the objective function in (\ref{max-min-sinr-relaxed}) is quasi-concave. Define $\Omega \triangleq \{\psi_{li}^{nl},X_{nl},Y_{nl}\}$ for $(l,i,n)\in [L]\x[3]\x[K]$, the set of optimization variables. The objective function of (\ref{max-min-sinr-relaxed}) is

\begin{align}
f(\Omega)=\min_{k,j} \frac{\Big| \sum_{i=1}^3 \psi_{ji}^{[kj]}\sqrt{ G_{ji}^{[kj]}} \lambda_{ji}^{[kj]}\Big|^2}{ X_{kj}^2 +  Y_{kj}^2 + \sigma^2_f}. \notag
\end{align}

For every $\gamma \geq 0$, the upper-level set of $f(\Omega)$ is

\begin{align}
U(f,\gamma)&=\{\Omega: f(\Omega) \geq \gamma\} \notag \\
&=\{\Omega: \frac{\Big| \sum_{i=1}^3 \psi_{ji}^{[kj]}\sqrt{ G_{ji}^{[kj]}} \lambda_{ji}^{[kj]}\Big|^2}{ X_{kj}^2 +  Y_{kj}^2 + \sigma^2_f} \geq \gamma, \forall (k,j)\} \notag \\
&= \{\Omega:||V_{kj}|| \leq \frac{1}{\sqrt{\gamma}} \sum_{i=1}^3 \psi_{ji}^{[kj]}\sqrt{ G_{ji}^{[kj]}} \lambda_{ji}^{[kj]}, \forall (k,j)\}, \notag
\end{align}
where $V_{kj} \triangleq [X_{kj},Y_{kj},1]$. Because $U(f,\gamma)$ can be represented as a second order cone, it is a convex set. Therefore, $f(\Omega)$ is quasi-concave.

\end{appendices}

% **************************************************************************************************
% References
% **************************************************************************************************
\bibliographystyle{IEEEtran}
\bibliography{directional}

\end{document}